\documentclass[sn-mathphys-num]{sn-jnl}

\usepackage{graphicx}%
\usepackage{multirow}%
\usepackage{amsmath,amssymb,amsfonts}%
\usepackage{amsthm}%
\usepackage{mathrsfs}%
\usepackage[title]{appendix}%
\usepackage{xcolor}%
\usepackage{textcomp}%
\usepackage{manyfoot}%
\usepackage{booktabs}%
\usepackage{algorithm}%
\usepackage{algorithmicx}%
\usepackage{algpseudocode}%
\usepackage{listings}%

\theoremstyle{thmstyleone}%
\theoremstyle{thmstyletwo}%

\theoremstyle{thmstylethree}%

\begin{document}

\title[Article Title]{Quantum phase transitions in one-dimensional nanostructures: a comparison between DFT and DMRG methodologies}


\author[1]{\fnm{T.} \sur{Pauletti}}\email{tatiana.pauletti@unesp.br}

\author[1]{\fnm{M.} \sur{Sanino}}\email{marina.sanino@unesp.br}

\author[1]{\fnm{L.} \sur{Gimenes}}\email{lucas.gimenes@unesp.br}

\author*[1]{\fnm{I. M.} \sur{Carvalho}}\email{isaac.carvalho@unesp.br}

\author[1]{\fnm{V. V.} \sur{França}}\email{vivian.franca@unesp.br}

\affil*[1]{\orgdiv{Institute of Chemistry}, \orgname{São Paulo State University}, \orgaddress{\street{Francisco Degni 55}, \city{Araraquara}, \postcode{14800-090}, \state{São Paulo}, \country{Brazil}}}




\abstract{\textbf{Context:} In the realm of quantum chemistry, the accurate prediction of electronic structure and properties of nanostructures remains a formidable challenge. Density Functional Theory (DFT) and Density Matrix Renormalization Group (DMRG) have emerged as two powerful computational methods for addressing electronic correlation effects in diverse molecular systems. We compare ground-state energies ($e_0$), density profiles ($n$) and average entanglement entropies ($\bar S$) in metals, insulators and at the transition from metal to insulator, in homogeneous, superlattices and harmonically confined chains described by the fermionic one-dimensional Hubbard model. While for the homogeneous systems there is a clear hierarchy between the deviations,  $D\%(\bar S)<D\%(e_0)< \bar D\%(n)$, and all the deviations decrease with the chain size; for superlattices and harmonical confinement the relation among the deviations is less trivial and strongly dependent on the superlattice structure and the confinement strength considered. For the superlattices, in general increasing the number of impurities in the unit cell represents less precision on the DFT calculations. For the confined chains, DFT performs better for metallic phases, while the highest deviations appear for the Mott and band-insulator phases. This work provides a comprehensive comparative analysis of these methodologies, shedding light on their respective strengths, limitations, and applications.

\textbf{Methods:} The DFT calculations were performed using the standard Kohn-Sham scheme within the BALDA approach. It integrated the numerical Bethe-Ansatz (BA) solution of the Hubbard model as the homogeneous density functional within a local-density approximation (LDA) for the exchange-correlation energy. The DMRG algorithms were implemented using the ITensor library, which is based on the Matrix Product States (MPS) ansatz. The calculations were performed until the energy reaches convergence of at least $10^{-8}$.}




\maketitle

\section{Introduction}\label{sec1}

Reliable predictions of nanostructures properties are essential for the development of advanced quantum technologies. Density Functional Theory (DFT)~\cite{geerlings2003conceptual,cohen2012challenges} and Density Matrix Renormalization Group (DMRG)~\cite{dmrg2} methods stand out prominently in this pursuit.  DMRG has demonstrated remarkable success in capturing strong correlation effects by systematically optimizing the wavefunction in a reduced Hilbert space, but gets computationally expensive with the size of the system.

On the other hand, DFT offers a cost-effective approach to study large systems and is widely used as an electronic structure method. However, the inherent approximations in exchange-correlation functionals can lead to inaccuracies, particularly in strongly correlated systems. Despite of that Kohn-Sham DFT \cite{PhysRev.140.A1133,dft1} has been demonstrated to be reliable to describe for example {\it i)} transport properties in the weakly coupled repulsive regime \cite{ref1_vivian}, {\it ii)} phases in cold trapped atoms \cite{ref2_vivian, ref3_vivian}, {\it iii)} charge gap \cite{ref5_vivian}, and {\it iv)} electrical response properties of homogeneous metals \cite{ref4_vivian}. 

In this work, we provide a detailed comparative assessment of DFT and DMRG in nanostructures described by the one-dimensional Hubbard model. We explore homogeneous, superlattices and harmonically confined chains, addressing key properties such as ground-state energies ($e_0$), density profiles ($n$) and average entanglement entropies ($\bar S$), the latter being commonly used to detect and characterize quantum phase transitions.  While for the homogeneous systems one finds a clear hierarchy between the deviations,  $D\%(\bar S)<D\%(e_0)< \bar D\%(n)$, for superlattices and harmonically confined chains the relation among deviations is non-trivial and dependent on the superlattice structure and the confinement strength adopted. For the superlattices, in general increasing the number of impurities in the unit cell makes DFT calculations less precise. For the confined systems, DFT performs better for metallic phases, while the highest deviations appear for the Mott and band-insulator phases. 

\section{Model and Methods}\label{sec2}

We consider one-dimensional (1D) systems described by the fermionic Hubbard Hamiltonian,
\begin{eqnarray}
\hat{H} = -t\sum_{i,\sigma}(\hat{c}^{\dagger}_{i,\sigma}\hat{c}_{i+1, \sigma}+\hat{c}_{i,\sigma}\hat{c}^{\dagger}_{i+1, \sigma})
+U\sum_i\hat{n}_{i,\uparrow}\hat{n}_{i,\downarrow}+\sum_{i,
+-\sigma}V_i\hat{n}_{i,\sigma},
  \label{hamiltonian}
\end{eqnarray}
where $\hat{c}^{(\dagger)}_{i,\sigma}$ annihilates (creates) an electron with spin $\sigma=\uparrow,\downarrow$ at site $i$, and $\hat{n}_{i,\sigma} = \hat{c}^{\dagger}_{i,\sigma}\hat{c}_{i,\sigma}$ are number operators. The model considers nearest-neighbour hopping, on-site interactions and local external potential terms with coefficients $t$, $U$, and $V_{i}$, respectively. The ground-state is calculated by fixing the total number of electrons $N$ in a chain of size $L$ and the total null magnetization, such that $\sum_{i} \left\langle \hat{n}_{i,\uparrow} \right\rangle = \sum_{i}\left\langle \hat{n}_{i,\downarrow} \right\rangle = N/2$. We consider different average densities $n=N/L$, restricted to $0\leq n \leq 2$ (single energy band)  and  impose open boundary conditions at the chain ends. Throughout this work, we set $t=1$, which defines the unit of energy. 

\begin{figure}[t!]
\begin{centering}
\includegraphics[scale=0.6]{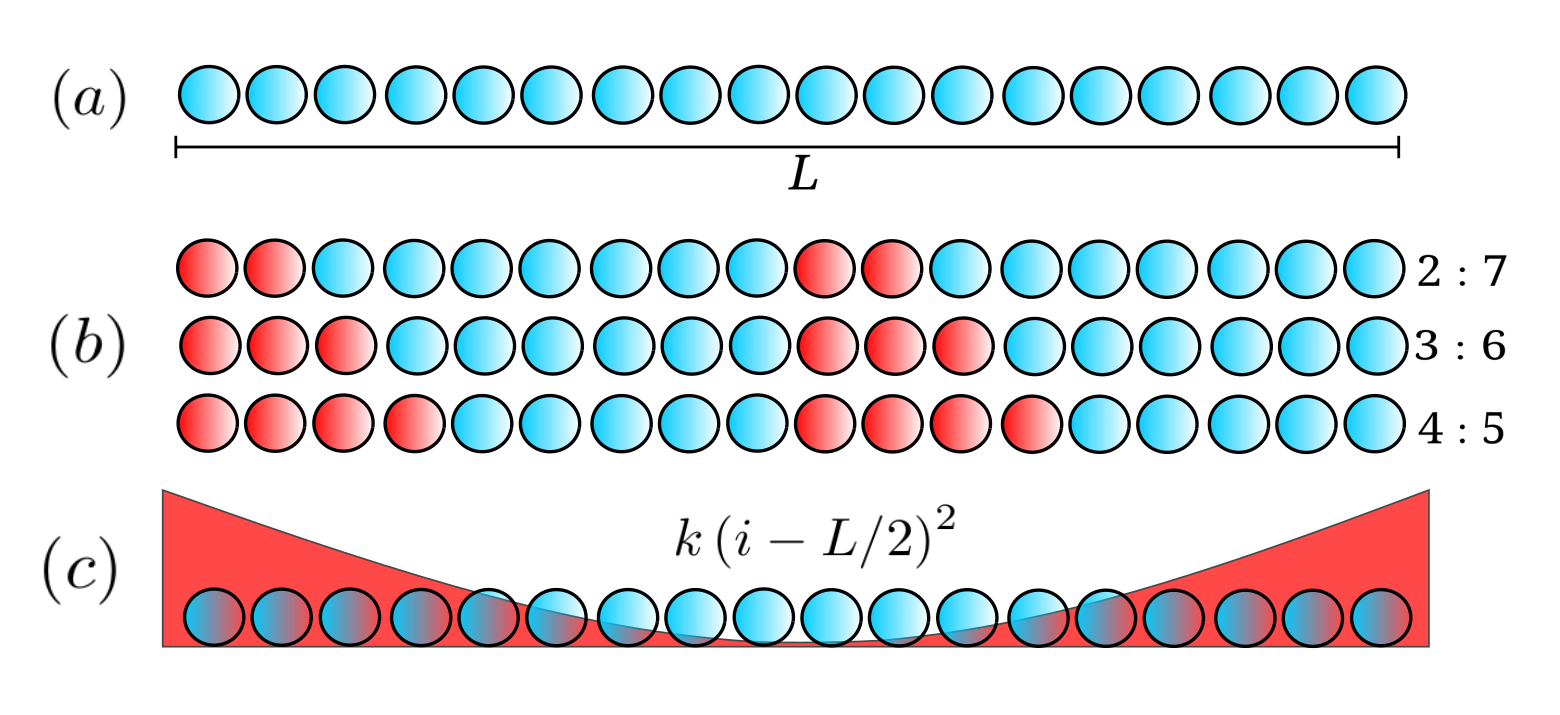}
\par\end{centering}
\caption{Schematic setup of (a) a finite homogeneous lattice with $V_i=0$ (blue circles), (b) superlattices with distinct modulations of the periodic $V_i\neq 0$ (red circles), for a fixed unit cell of 9 sites ($2:7$, $3:6$ and $4:5$) and (c) a chain under a harmonic confinement centered at $L/2$, $V_i=k{(i-L/2)}^{2}$.\label{fig1}}
\end{figure}

At the limit $V_i = 0$ the Hamiltonian of Eq.~(\ref{hamiltonian}) reduces to the standard Hubbard model. In this case, for $n<1$ the ground-state represents a metallic phase, while for the half-filling ($n=1$) the system consists of a Mott insulating phase for arbitrarily small $U>0$~\cite{giamarchi2003quantum}. When $V_i$ is periodically modulated, thus simulating superlattices (as in Fig. \ref{fig1}b), the potential may induce a quantum phase transition from metal to insulating \cite{PhysRevB.87.214407, PhysRevB.58.9607,PhysRevB.102.235151}. Similarly the Mott metal-insulator transition may be induced by the harmonical potential $V_i=k(i-L/2)^2$ (as in Fig. \ref{fig1}c), widely used in state-of-the-art cold atoms experiments.

In this paper, we address the performance of DFT $-$ when compared to numerically exact DMRG calculations $-$ in describing the properties of the Hubbard model for all the three 1D systems: homogenous, superlattices and harmonically confined chains, as schematically illustrated in Figure \ref{fig1}. 

We focus on three main quantities: the per-site ground-state energy $e_0$, the density profile $\{n_i\}$ and its respective nonlocal correlations, quantified by the average single-site entanglement 
\begin{eqnarray}
    \bar S=\frac{1}{L}\sum_{i}^{L}S_i,
\end{eqnarray}
with
\begin{eqnarray}
    S_i=-{\rm Tr}[ {\rho}_i {\rm log}_2 {\rho}_i]=-\sum_{k} {\rm w}_{i,k}{\rm log}_2 {\rm w}_{i,k},
\end{eqnarray}
where ${\rm w}{_{i,k}}$ are the eigenvalues of the $i$-site reduced density matrix ${\rho}_{i}={\rm Tr}_{L-1} [\rho_{GS}]$ calculated by tracing out the remaining $L-1$ sites from the ground-state density matrix $\rho_{GS}$. This type of entanglement has been proved to be relevant in analyzing the localization and itinerancy of the indistinguishable particles~\cite{zanardi,tichy2013entanglement,canellaVV}. In the site-occupation basis the reduced Hilbert space has dimension $d=4$, and thus ${\rm w}_{i,k}$ are the occupation probabilities with $k=\uparrow, \downarrow, 2,0$, such that $\sum_k {\rm w}_{i,k}=1$. The set $\{{\rm w}_{i,k}\}$ can be calculated by firstly computing the paired probability $ {\rm w}_{i,2}= \langle \hat n_{i\uparrow}\hat n_{i\downarrow}\rangle=\partial e_0/\partial U$. From this term one obtain the remaining probabilities: the unpaired, $ {\rm w}_{i,\uparrow}= {\rm w}_{i,\downarrow}=\left\langle \hat{n}_i \right\rangle/2- {\rm w}_{i,2}$, and the empty probability ${\rm w}_{i,0}=1- {\rm w}_{i,\uparrow}- {\rm w}_{i,\downarrow}- {\rm w}_{i,2}$. The DMRG algorithms were implemented using the ITensor library  \cite{itensor}. This library is based on the Matrix Product States (MPS) ansatz. In this context, up to $800$ sweeps were utilized to reach an energy convergence of at least $10^{-8}$. The precision of the MPS representations was essentially controlled by configuring the bond link, which its maximum is setting to approximately $3000$.

Via DFT, the per-site ground-state energy $e_0$ and the density profile $\{n_i\}$ are obtained via standard Kohn-Sham scheme \cite{PhysRev.140.A1133,dft1} within BALDA approach: in which the numerical Bethe-Ansatz (BA) solution of the Hubbard model is considered as the homogeneous density functional within a local-density approximation (LDA) for the exchange-correlation energy (for a review see \cite{CAPELLE201391}).

Now for the entanglement, which requires the energy derivative with respect to $U$, ${\rm w}_2=\partial e_0/\partial U$, 
to avoid errors related to the numerical derivative of the BALDA approach described above, we adopt instead the analytical derivative of the FVC parameterization for the energy ~\cite{FVC}, which (for non-magnetized systems) is given by 

\begin{eqnarray}
{\rm w}_2^{FVC}(n,U)=
      \frac{2n} {\beta(n,U)}\frac{\partial\beta(n,U)}{\partial U} \cos\left(\frac{\pi n}{\beta(n,U)}\right)
       -\frac{2}{\pi} \frac{\partial\beta(n,U)}{\partial U}\sin\left(\frac{\pi n}{\beta(n,U)}\right),\label{fvc}
\end{eqnarray}
where $\beta(n,U)=\beta(U)^{\sqrt[3]{U}/8}$ and
the function $\beta(U)$ is determined from
\begin{equation}
    \frac{\beta\left(U\right)}{\pi}\sin\left(\frac{\pi}{\beta\left(U\right)}\right)=2\int_{0}^{\infty}{\rm d}x\frac{J_{0}\left(x\right)J_{1}\left(x\right)}{x\left(1+e^{Ux/2}\right)},\label{lsco_eq2}
\end{equation}
with $J_0$ and $J_1$ the zero and first order Bessel functions, respectively. This analytical parametrization becomes exact by construction for {\it i)} $U\rightarrow 0$ (for which $\beta=2$), {\it ii)} $U\rightarrow\infty$ ($\beta = 1$), and {\it iii)} $n=1$ and any $U$ ($0\leqslant\beta\leqslant1$). For other $(n,U)$ regimes the FVC parametrization provides a reasonable approximation to the full Beth-Ansatz~\cite{PhysRevLett.20.1445,essler2005one} solution. 

\begin{figure}[t!]
\begin{centering}
\vspace{-0.6cm}
\includegraphics[scale=0.4]{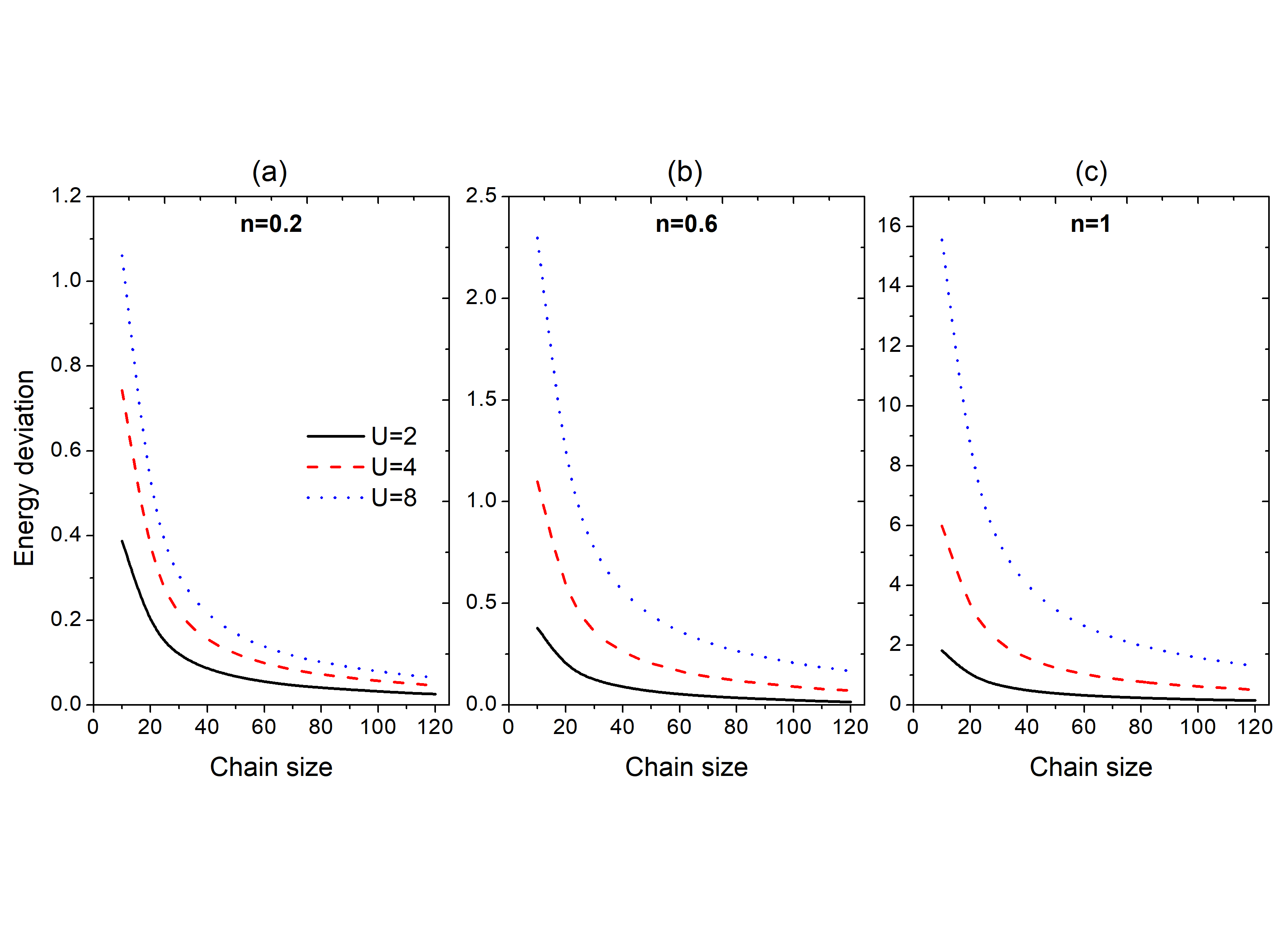}
\vspace{-1.4cm}
\par\end{centering}
\caption{DMRG-DFT deviation for the ground-state energy, Eq.(\ref{d_e}), as a function of the chain length for a finite homogeneous system for low density (a), intermediate density (b), and at half-filling (c), for several strengths of interaction $U$. \label{fig2}}
\end{figure}

\begin{figure}[h!]
\begin{centering}
\includegraphics[scale=0.4]{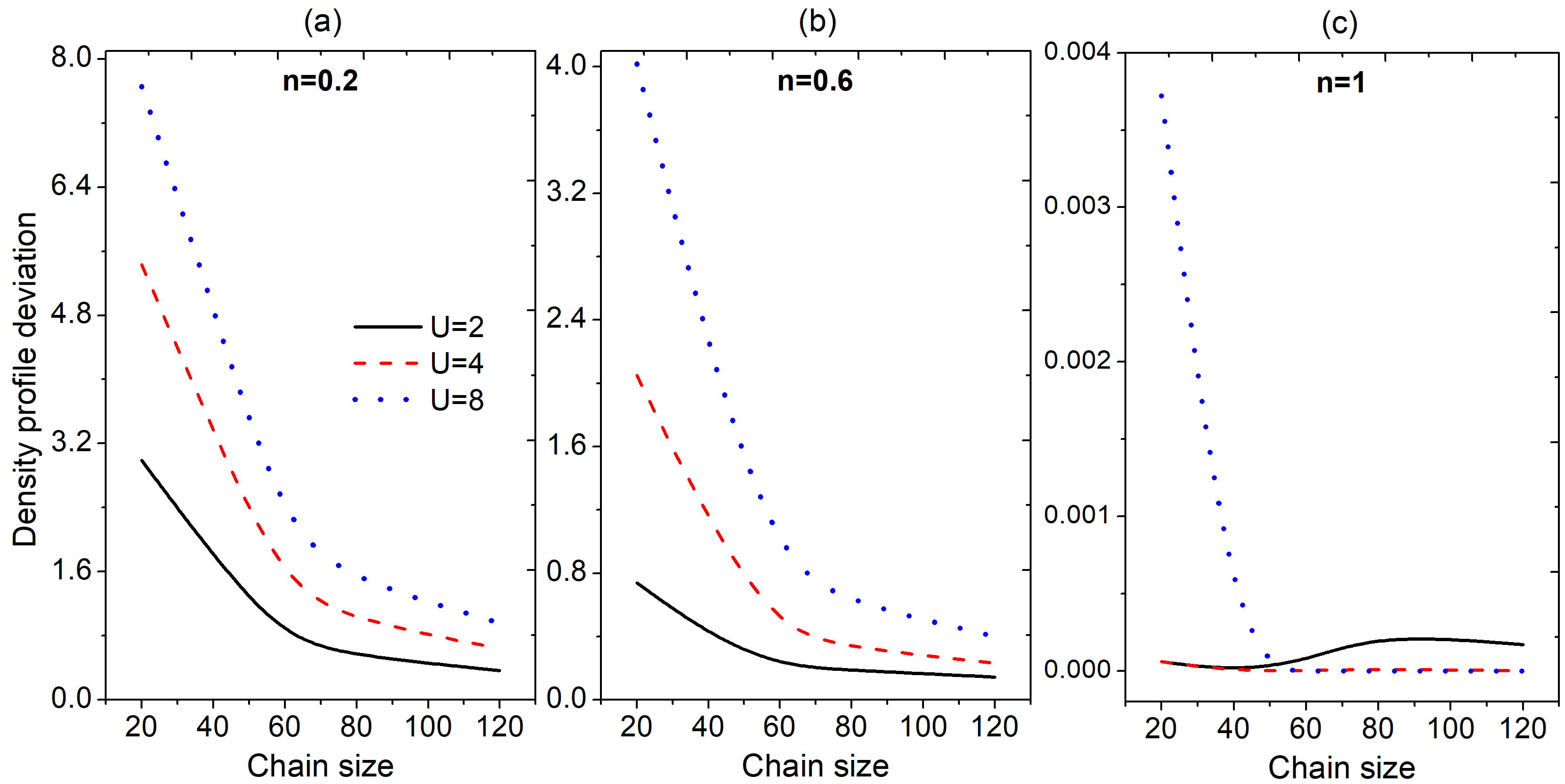}
\par\end{centering}
\caption{Mean percentage deviation of the local densities, Eq.(\ref{d_n}), as a function of the chain length for a homogeneous system at (a) $n=0.2$, (b) $n=0.6$, and (c) $n=1$ for several values of $U$. \label{fig3}}
\end{figure}

We thus adopt Eq.(\ref{fvc}) as approximation to the homogeneous chain and hence perform a LDA to obtain the double occupancy for the inhomogeneous systems: 

\begin{equation}
{\rm w}_{i,2}\approx {\rm w}_{i,2}^{LDA}={\rm w}_{2}^{FVC}(n,U)|_{n\rightarrow n_i},
\end{equation}
by using the density profile $\{n_i\}$ obtained via BALDA. 

The percentage deviations between DFT and DMRG calculations are then, for the energy

\begin{equation}
   D\%(e_0)= \left|\frac{e_{0}^{\rm DMRG}-e_{0}^{\rm DFT}}{e_{0}^{\rm DMRG}}\right|\times 100,\label{d_e}
\end{equation}
for the average single-site entanglement,

\begin{equation}
     {D}\%(\bar S)=\left|\frac{\bar S^{\rm DMRG}- \bar S^{\rm DFT}}{\bar S^{\rm DMRG}}\right|\times 100,\label{d_S}\end{equation}
while for the density profiles we quantify a mean percentage deviation, defined as:

\begin{equation}
     \bar {D}\%(n)=\frac{1}{L}\sum_i^L\left|\frac{n_{i}^{\rm DMRG}-n_{i}^{\rm DFT}}{n_{i}^{\rm DMRG}}\right|\times100.\label{d_n}
\end{equation}

\section{Results and Discussion}\label{sec3}

\subsection{DFT x DMRG: energy, density profile and entanglement}

First, we analyze the DFT performance in reproducing the DMRG results for finite chains. In Figure \ref{fig2} we present the percentage deviation for the per-site ground-state energy. We see that in general DFT performs better for low densities and weak interactions, in both cases due to less influence of the electron-electron correlations. We find that the error monotonically decays by increasing the chain length $L$ for all $n,U$, since our DFT-LDA approach becomes exact in the limit of $L\rightarrow \infty$. 

\begin{figure}[tbh]
\begin{centering}
\includegraphics[scale=0.4]{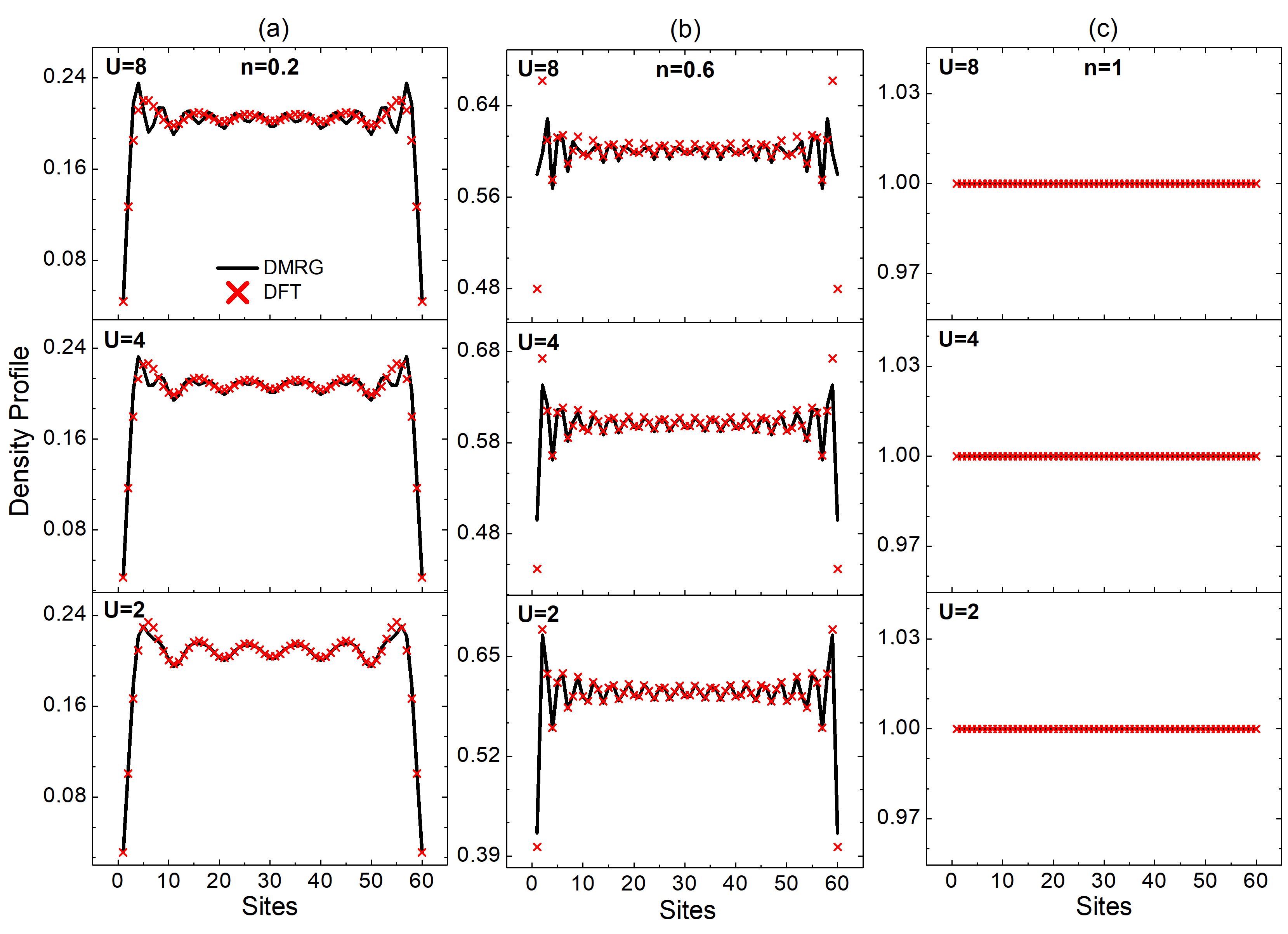}
\par\end{centering}
\caption{Density profile for the homogeneous system with $L=60$ sites, for several interactions $U$ and for distinct average filling factors: (a) $n=0.2$, (b) $n=0.6$, and (c) $n=1$. \label{fig4}}
\end{figure}

 Now considering the DFT performance for the local densities $\{n_i\}$, we show in Figure \ref{fig3}  the mean deviation defined in Eq.(\ref{d_n}). For the Mott insulator regime, $n=1$ and any $U>0$, DFT is very precise with average deviations smaller than $0.004\%$ due to the suppression of the Friedel-type oscillations~\cite{PhysRevB.58.10225,PhysRevB.79.195114}. For the metallic regime, $n<1$, the open ends of the chain act as effective impurities, inducing Friedel oscillations in the density distribution, which are not completely described by DFT, as can be confirmed by Figure \ref{fig4}. Since the Friedel oscillations are more pronounced for low densities, the deviations is greater for $n=0.2$ (Fig. \ref{fig3}a). In contrast, the smallest deviation for the energy is precisely for the case of $n=0.2$ (Fig. \ref{fig2}a). We attribute this to error cancellations: while Eq.(\ref{d_n}) sum up the local errors, the under and over DFT estimates for the density profile leads to error cancellation for the total energy. Nevertheless, for chain sizes $L\gtrsim60$ DFT density profiles are also reliable, with mean deviations smaller than $2\%$.
 
\begin{figure}[tbh]
\begin{centering}
\includegraphics[scale=0.4]{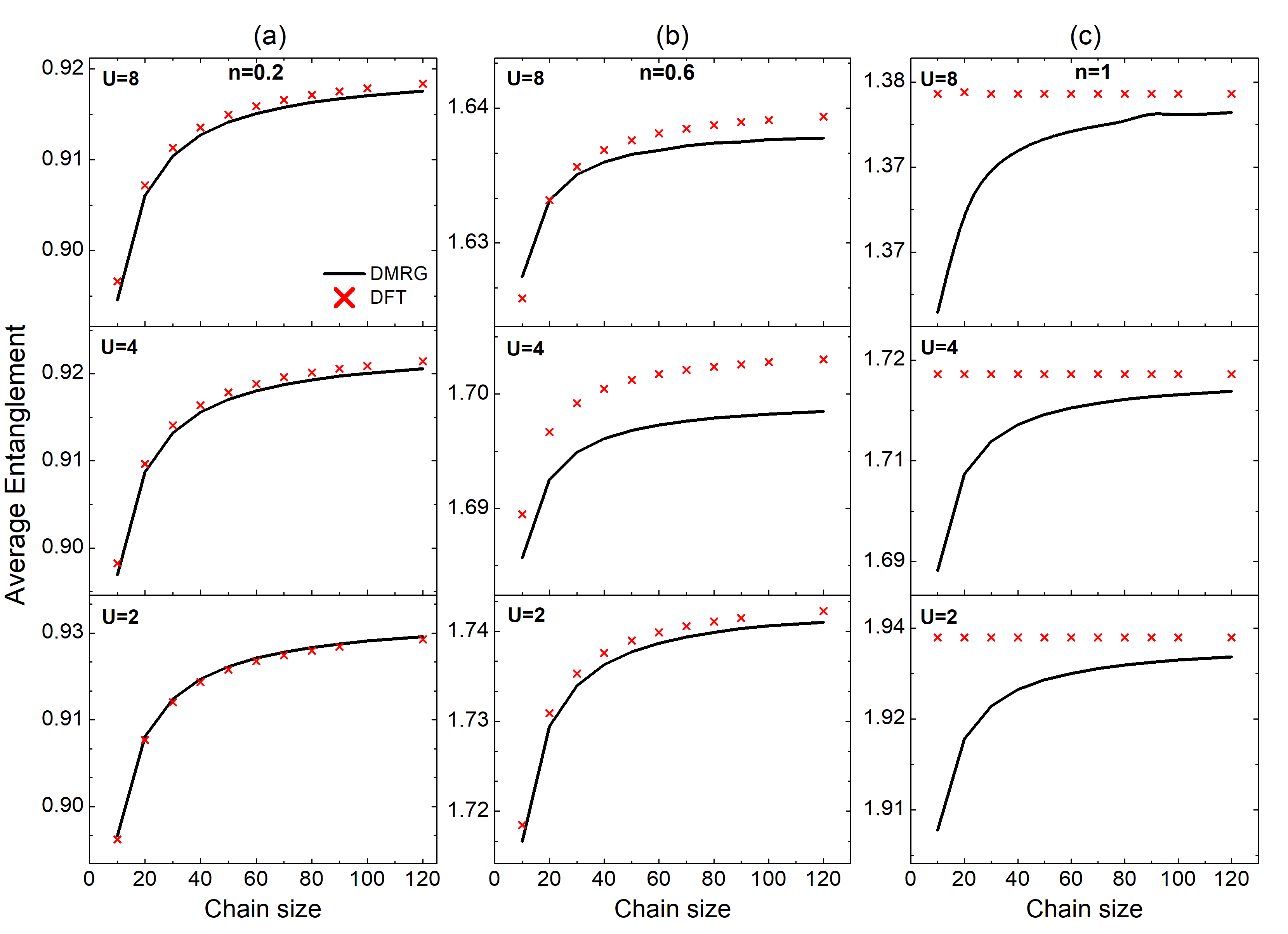}
\par\end{centering}
\caption{Average entanglement $\bar S$ as a function of the chain length of homogeneous systems for several interaction regimes and different densities: $n=0.2$ (a), $n=0.6$ (b), and $n=1$ (c).
\label{fig5}}
\end{figure}

For the average single-site entanglement, in Figure \ref{fig5}, one finds a good performance of DFT calculations even for the worse regimes of parameters (for $L=10$, $n=1$, and $U=4$): the maximum deviation observed is $1.73\%$. This very good performance of DFT for the average entanglement also comes from error cancellations: the von Neumann entropy is a density functional \cite{zanardi,tichy2013entanglement}, thus it reflects the DFT difficulty of reproducing the Friedel oscillations by under and overestimating the single-site entanglement along the chain, hence leading to a good average. As $\bar S$ has been widely used to detect and characterize quantum phase transitions \cite{canellaVV,PhysRevB.101.214522,PhysRevA.87.032311,PhysRevA.86.033622,PhysRevA.81.052321,PhysRevA.73.042320,PhysRevB.73.085113,PhysRevA.66.032110,PhysRevB.108.184302}, our results then show that DFT represents a reliable and powerful method to be used in Hubbard chains, as we explore below in inhomogeneous chains.

\subsection{DFT x DMRG: detecting quantum phase transitions}

\begin{figure}[tbh]
\begin{centering}
\includegraphics[scale=0.4]{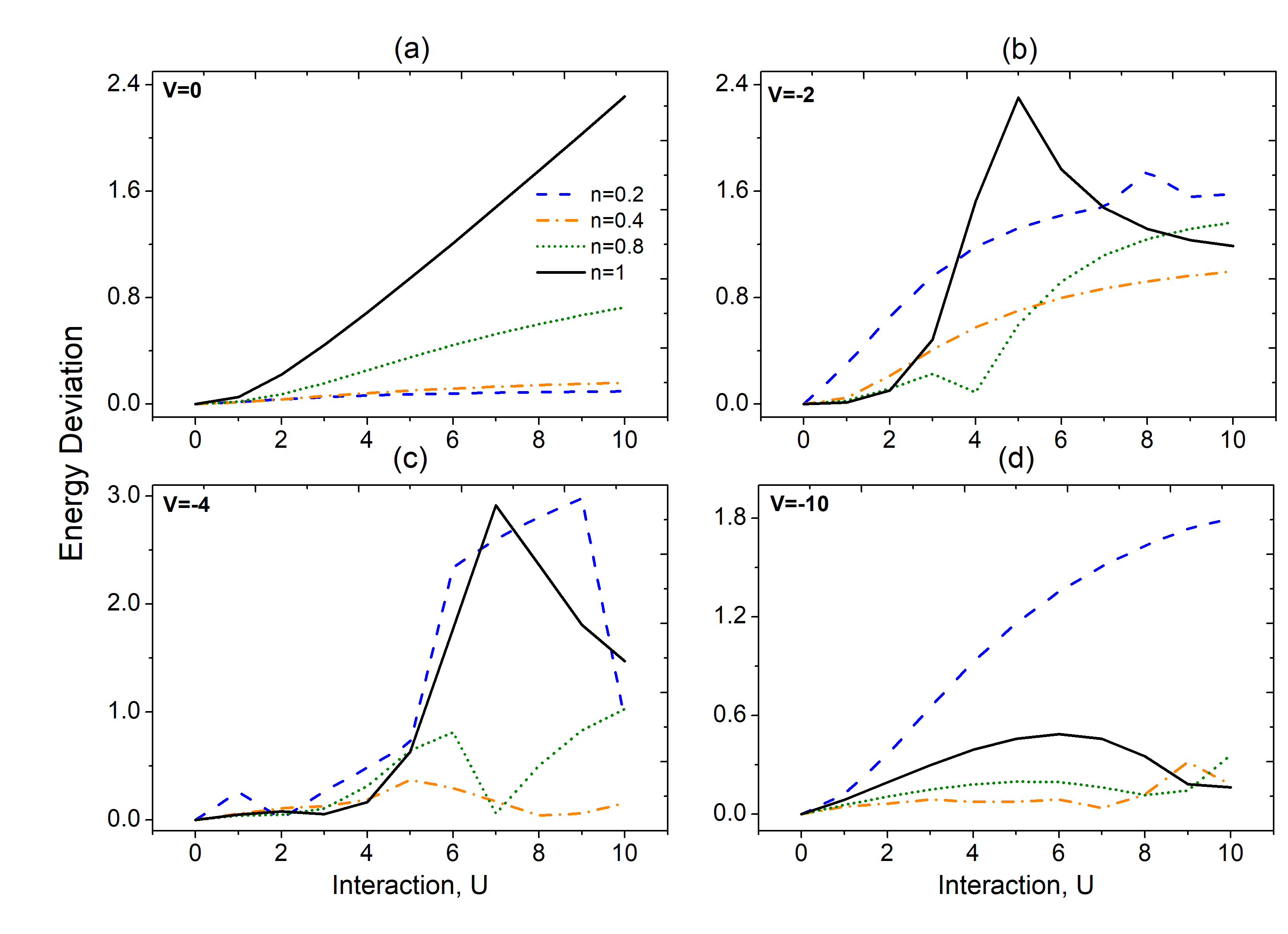}
\par\end{centering}
\caption{
DMRG-DFT deviation for the ground-state energy of the superlattice $2:7$ as a function of interaction $U$, for several periodic potential strengths: (a) $V=0$, (b) $V=-2$, (c) $V=-4$ and (d) $V=-10$. \label{fig6}}\end{figure}

We focus on the performance of DFT calculations for identifying the Mott metal-insulator transition in superlattices and harmonically confined chains. Starting with the superlattices (with unit cell SL $2:7$), Figure \ref{fig6} shows that DFT is considerably precise in reproducing the ground-state energy: deviations increase in general by increasing the interaction $U$ and the superlattice potential $V$, but are smaller than $\sim 3\%$ for all the regimes of parameters considered.

\begin{figure}[tbh]
\begin{centering}
\includegraphics[scale=0.4]{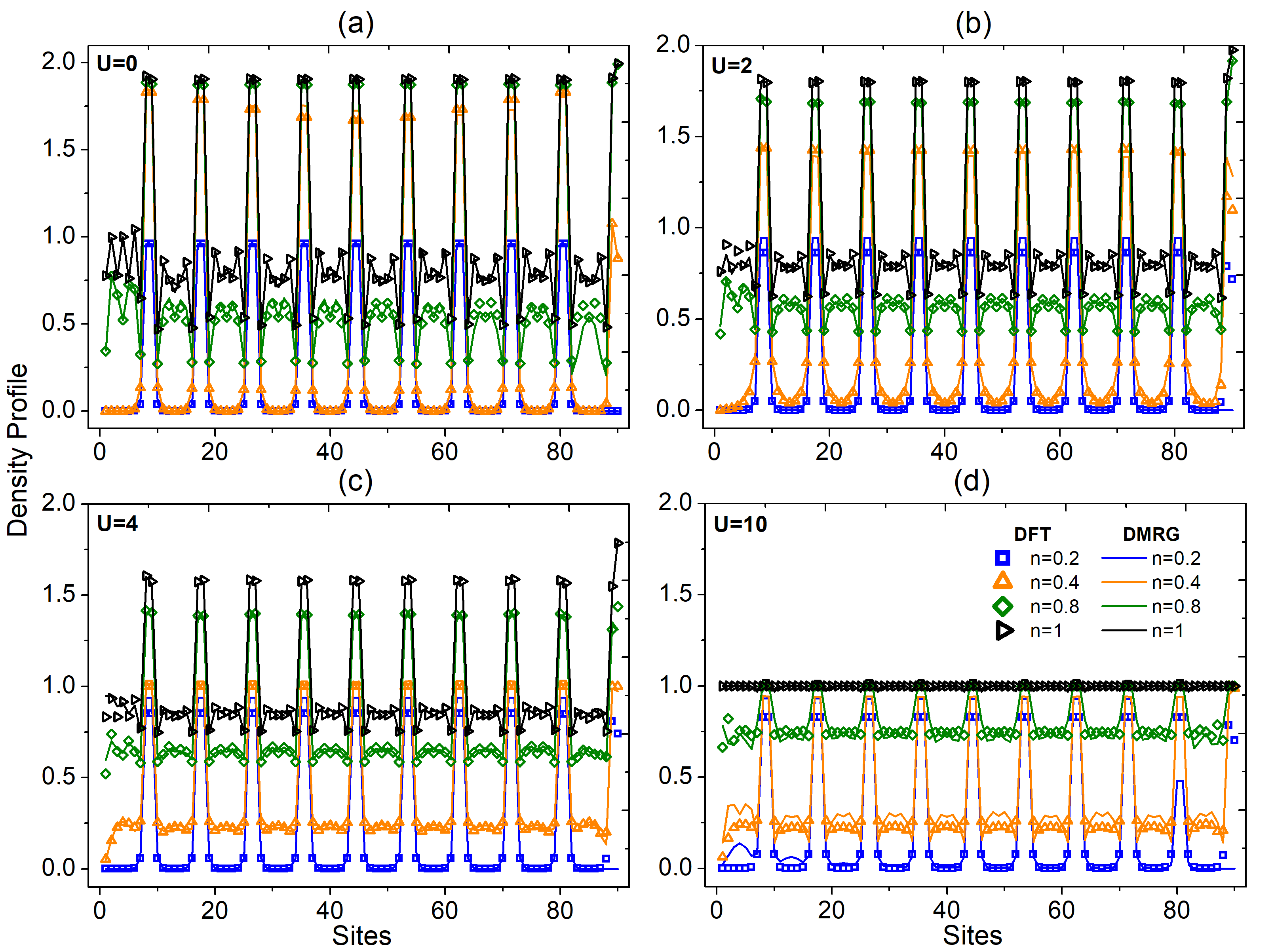}
\par\end{centering}
\caption{Density profile for a superlattice $2:7$ with $V=-4$, $L=90$ sites, for several densities and distinct interactions: (a) $U=0$, (b) $U=2$, (c) $U=4$ and (d) $U=10$.  \label{fig7}}
\end{figure}

 Accordingly, Figure \ref{fig7} reveals that also the density profile in the superlattice is reasonably well reproduced by DFT. For non-interacting systems, Fig. \ref{fig7}a, the density concentrates within the impurity sites, since the modulated potential is attractive, $V<0$. For $U=0$ and low average densities ($n=0.2$ and $n=0.4$), the particles essentially reside in the impurity sites, emptying the non-impurity ones. For higher average densities ($n=0.8$ and $n=1$), the impurity sites reach maximum occupation ($n_i=2$), thus the remaining particles are distributed also in the non-impurity sites. 

When turning the interaction on, Figs. \ref{fig7}b$-$d, we find that the non-impurity sites start to be filled with particles even for the low density regime $n=0.4$, once now the competition between $V<0$ and $U>0$ disfavors the maximum occupation at the impurity sites. Thus depending on the three parameters $V,n,U$ the interaction can be strong enough to avoid $n_i>1$ in the impurity sites, thus characterizing a local Mott insulator phase \cite{PhysRevB.58.9607,PhysRevB.87.214407}. This is seen already for $U\leq 2$ at $n=0.2$, while at $n=0.4$ this appears for $U\leq 4$ and at $n=0.8$ and $n=1$ only for $U\leq 10$, for this fixed $V=-4$.

\begin{figure}[tbh]
\begin{centering}
\includegraphics[scale=0.40]{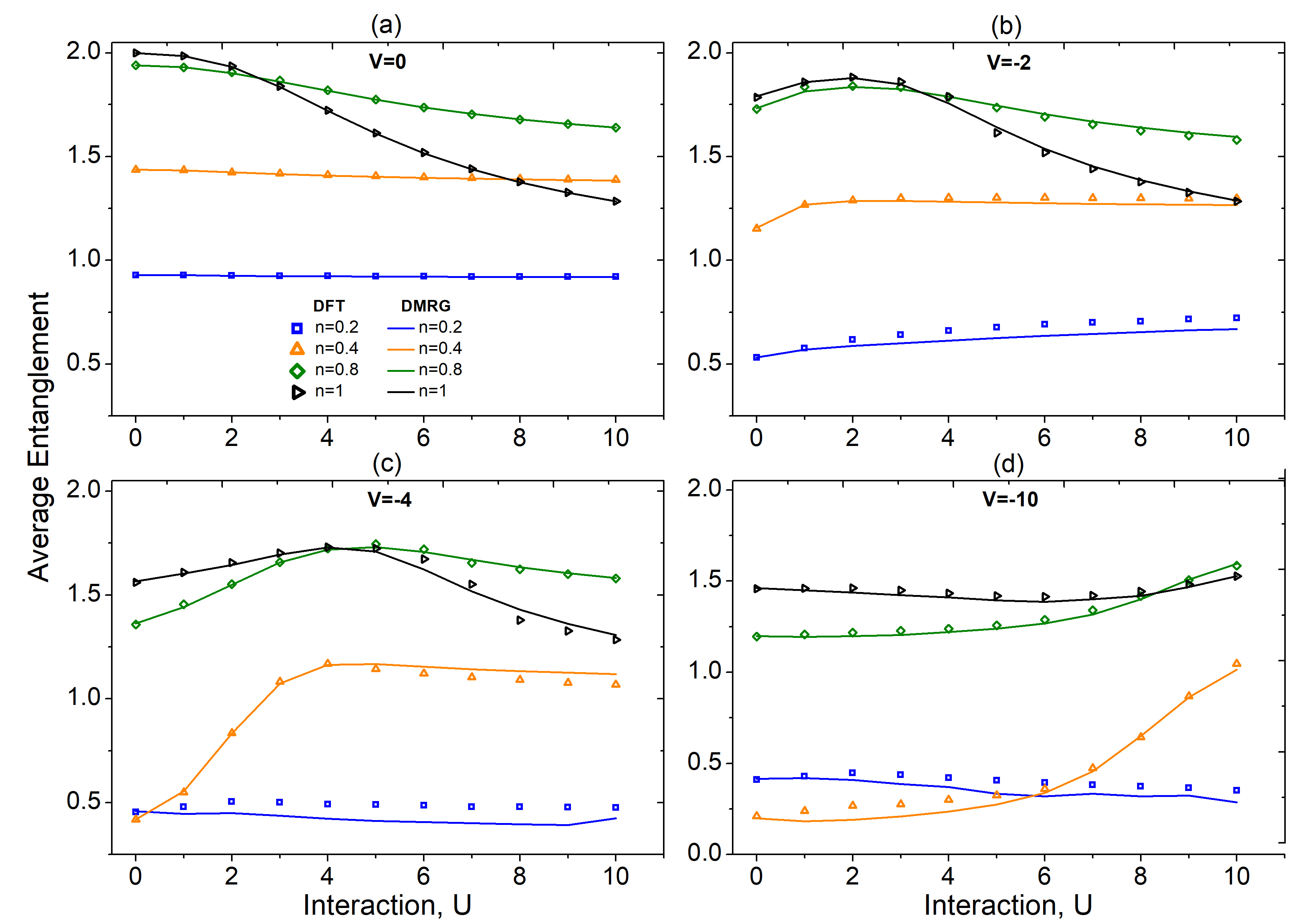}
\par\end{centering}
\caption{Average entanglement for a superlattice as a function of the on-site interaction, for several densities and distinct periodic potential strengths: (a) $V=0$, (b) $V=-2$, (c) $V=-4$ and (d) $V=-10$. }\label{fig8}
\end{figure}

We then analyze the average single-site entanglement as a function of $U$ in order to find signatures of the Mott metal-insulator transition within the superlattice. As shown in Figure \ref{fig8}, our results reveal that entanglement becomes non-monotonic with the density (crossing between distinct $n$  curves) whenever the entire chain or a portion of it reaches the Mott-insulator regime: at $V=0$ (homogeneous finite chain), it happens only for the $n=1$ curve, once this is the only regime for the Mott physics in the absence of the superlattice potential. In contrast, depending on  $V$, the non-monotonicity with $n$ is observed also for $n<1$, thus detecting the Mott insulator in a portion (the impurity sites) of the chain, as confirmed by Fig. \ref{fig7}. Consistently we also see that the interaction for which the curves cross increases by increasing $V$, supporting the interpretation that the mechanism behind the Mott insulator phase in superlattices is the competition between $U$ and $V$. Remarkably DFT properly captures all this physics.

\begin{figure}[tbh]
\begin{centering}
\includegraphics[scale=0.45]{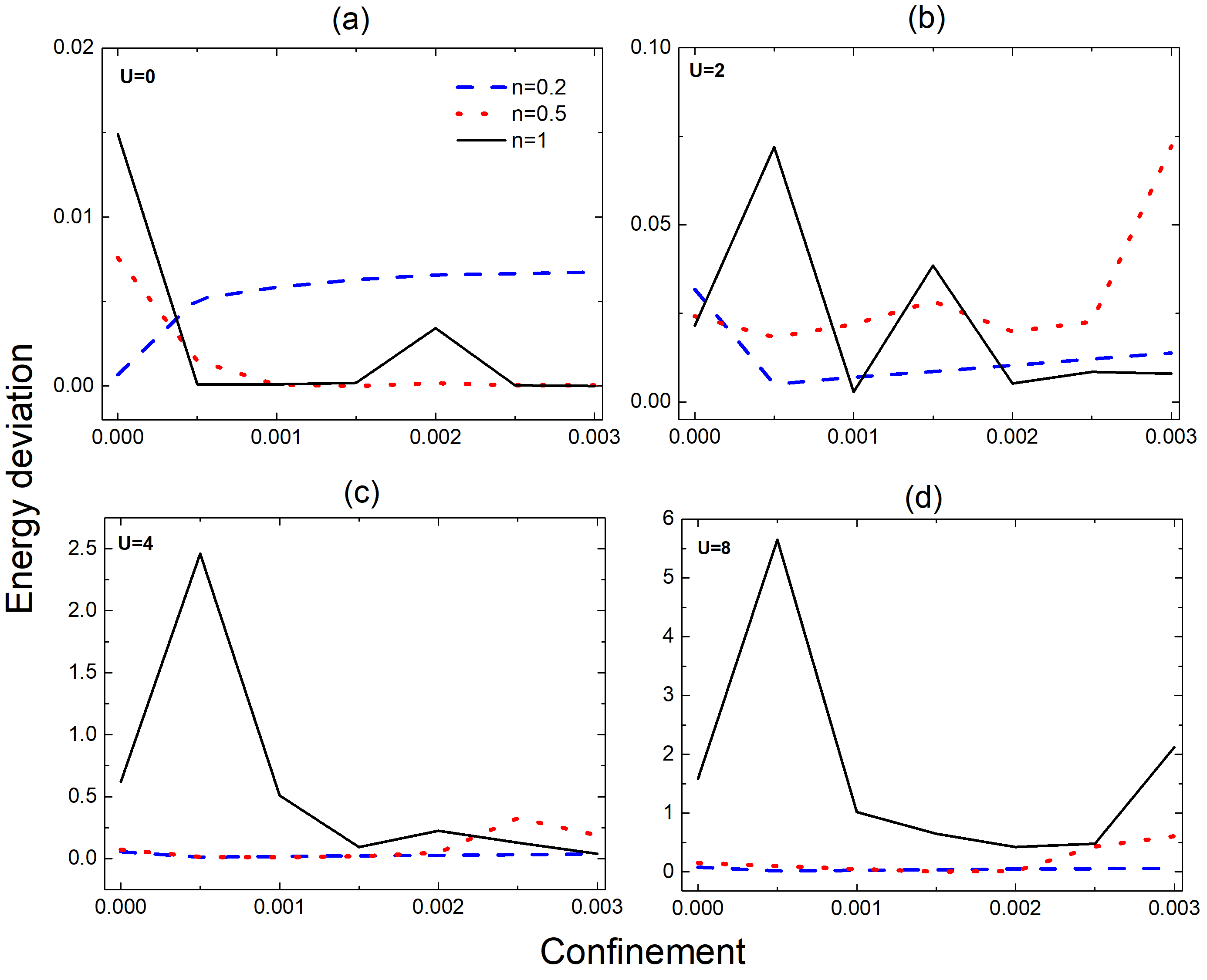}
\par\end{centering}
\caption{DMRG-DFT deviation for the ground-state energy as a function of the confinement curvature $k$ for several densities and interactions: (a) $U=0$, (b) $U=2$, (c) $U=4$ and (d) $U=8$.  \label{fig9}}
\end{figure}

In Figure \ref{fig9} we present the DFT performance in recovering the DMRG ground-state energies for harmonically confined chains ($V_i=k(i-L/2)^2$) as a function of the potential curvature $k$. Although the deviations increase with the interaction, we find that for typical strongly correlated systems ($U=4$) DFT energies deviate from DMRG by at most $2.5\%$, what can still be considered as a fair accuracy. 

\begin{figure}[tbh]
\begin{centering}
\includegraphics[scale=0.28]{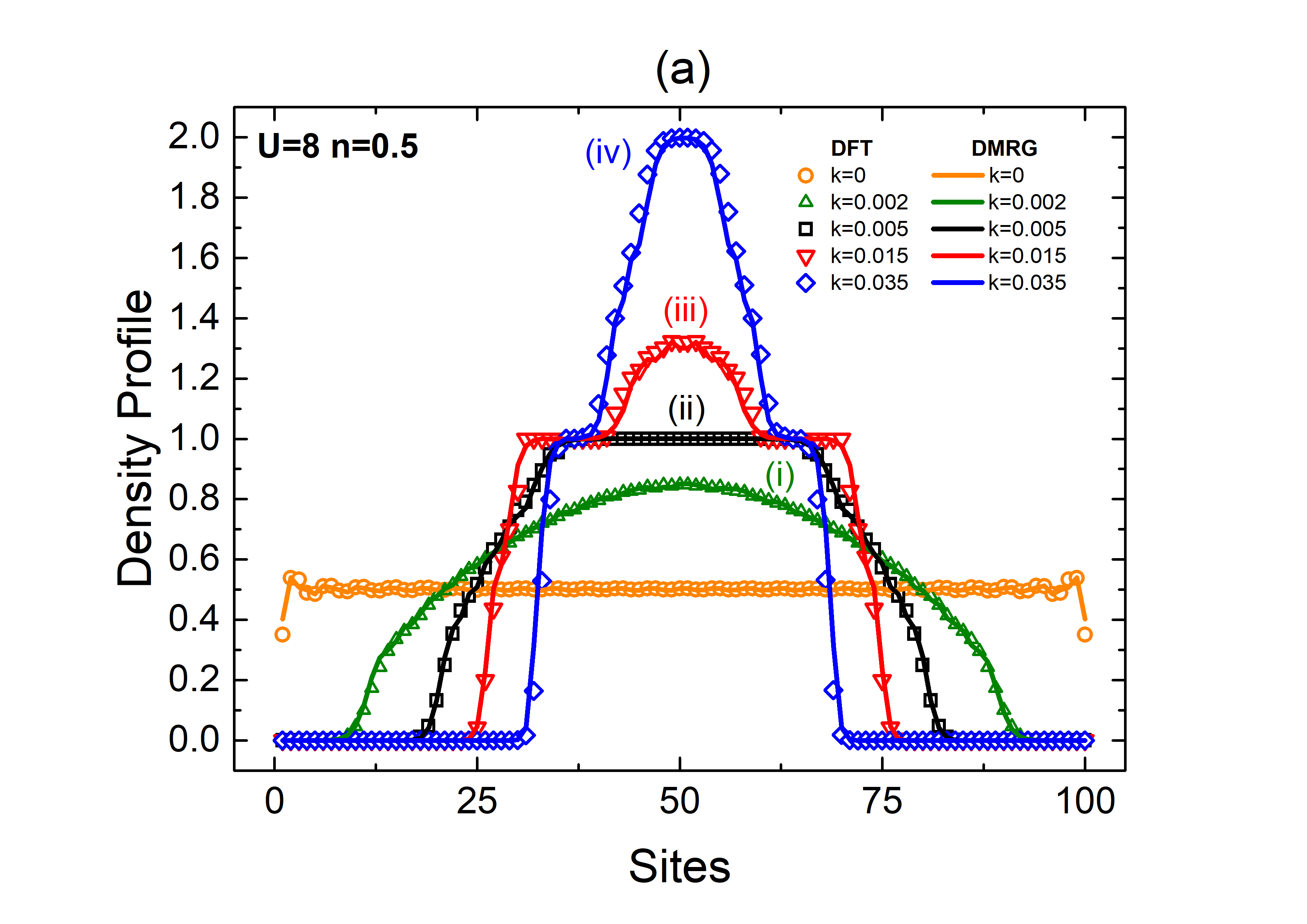}\hspace{-1cm}\includegraphics[scale=0.28]{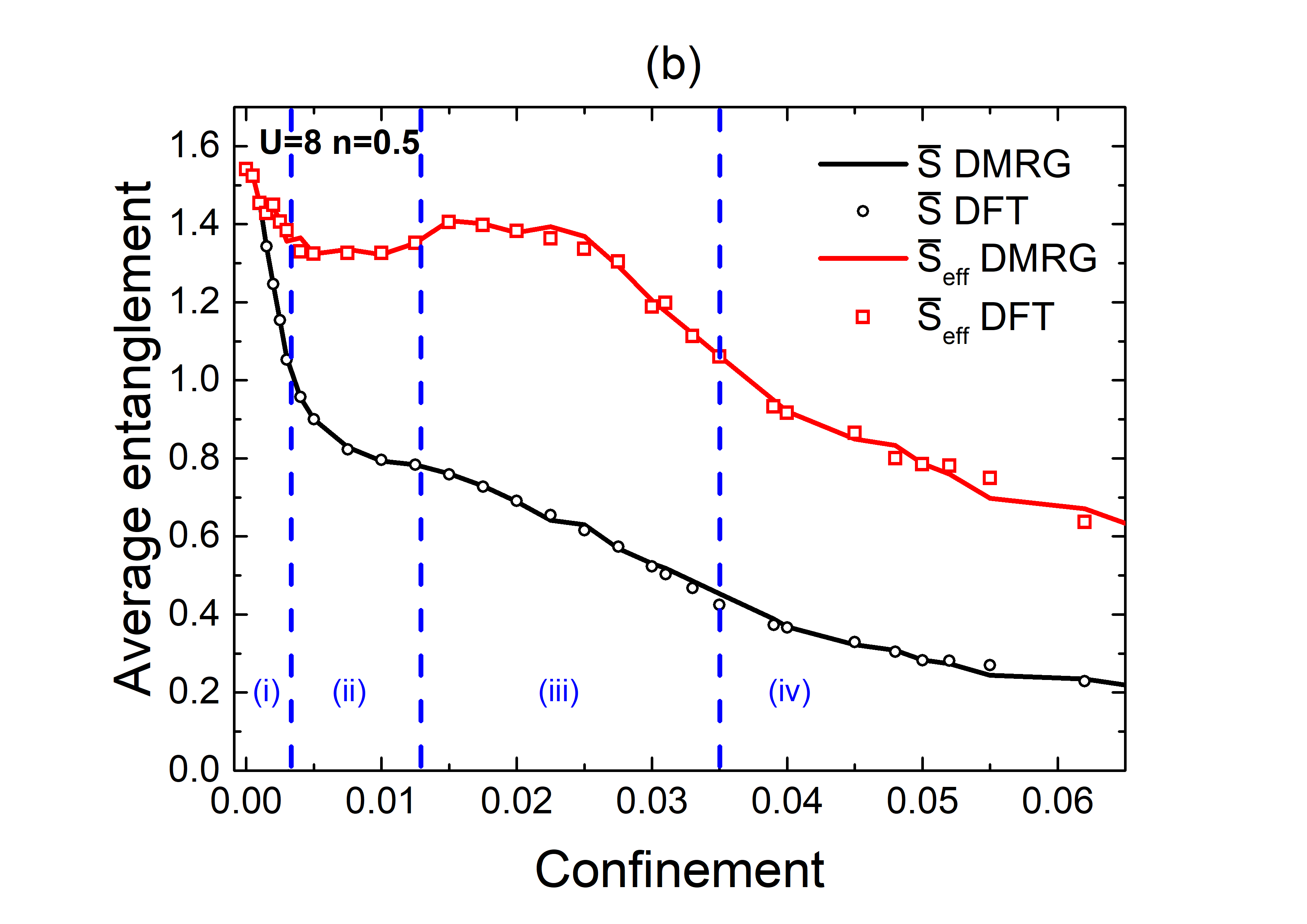}
\par\end{centering}
\caption{(a) Density profile for confined lattices, with $L=100$ sites, for several confinement curvatures $k$. (b) Entanglement as a function of the confinement curvature $k$: averaging over the entire chain, $\bar S$, and averaging only at the effective chain (central sites with occupation $n_i>0.001$), $\bar S_{eff}$. While in both cases DFT reproduces the DMRG data, one sees that $\bar S_{eff}$ is more sensitive to the transitions between the phases: (i) metallic, (ii) Mott-like insulating, (iii) metallic at the center of the chain, and (iv) band insulating at the core. \label{fig10}}
\end{figure}

For the density profile we find also a good agreement between the techniques, as can be seen in Figure \ref{fig10}a. We also observe that the effective chain $-$ defined by the central region where the particles are effectively distributed ($n_i>0.001$) $-$ decreases by increasing $k$, while the densities at the center of the chain increase. This occurs due to the strong potential at the wings. The density profile also reflects the metallic and insulating phases: (i) a metallic phase with $n_i<1$ at the entire chain, (ii) the Mott-like insulating phase, where the density at the center is kept fixed at $n_i=1$ and a stronger $V$ potential is required to produce $n_i>1$, (iii) another metallic phase with $n_i>1$ at the center of the chain; and (iv) a band-insulator phase, with maximum occupation $n_i=2$ at the core. 

As shows Figure \ref{fig10}b, the average single-site entanglement is very well reproduced by DFT calculations in harmonically confined chains. Notice that at $\bar S$ the signatures of the distinct phases are very subtle, therefore in Ref. \cite{PhysRevA.74.042325} the entanglement derivative was used to clearly identify the phases induced by the harmonical confinement. But numerical derivatives require a very large number of data to be precise, thus we here propose an alternative analysis to the transitions, which consists of considering the average single-site entanglement only at the effective chain, $\bar S_{eff}\equiv \bar S(n_i>0.001)$, i.e. for the sites that are effectively occupied. Fig. \ref{fig10}b reveals that $\bar S_{eff}$ signs better the distinct phases: it decreases monotonically with $k$  in the metallic phase (i);  has a local minimum at the Mott-like transition (ii); a local maximum at the metallic regime (iii); and a monotonic decreasing for the band-insulator phase (iv).

\begin{table}[tb]
\centering
\caption{Deviations for the ground-state energy (Eq.(\ref{d_e})), the average entanglement (Eq. (\ref{d_S})), and the density profile (Eq. (\ref{d_n})), for homogeneous chains ($U=8$, $n=0.6$) of distinct sizes $L$, superlattices ($U=8$, $n=0.4$, $V=-2$, $L=36$) with different structures SL, and harmonical confinement ($U=8$, $n=0.5$, $L=100$) for distinct curvatures $k$. }
    \begin{tabular}{|r|c|c|c|}
    \hline
        &{\bf GS energy}& {\bf density profile} & {\bf entanglement}\\
        &$D\%(e_0)$&$\bar D\%(n)$&  $D\%(\bar S)$\\
                \hline
        {\bf Homogeneous} & & & \\
        $L=10$ & $2.30$ & $3.09$ & $0.43$ \\
        $L=50$ & 0.44 & 0.88 & 0.14\\
        $L=100$ & 0.21 & 0.28 & 0.11\\
           \hline
        {\bf Superlattices} &   &&\\
        SL $2:7$ & 1.14 & 12.05 & 2.56 \\
        SL $3:6$ & 1.85 & 9.60 & 1.35\\
        SL $4:5$ & 1.51 & 17.19 & 2.92\\
          \hline
         {\bf Harmonical Confinement}& & & \\
         metal (i), $k=0.002$ & 0.01 & 7.94 & 0.16\\
         Mott-like insulating (ii), $k=0.005$ & 0.12 & 24.40 & 0.02\\
         metal (iii), $k=0.015$ &0.01 &3.96 &0.20 \\
         band insulator (iv), $k=0.035$ & 22.17 & 8.84 & 6.62\\
        \hline
         
    \end{tabular}
   
\end{table}

Finally in Table 1 we present the deviations for a few representative cases of each of the systems, homogeneous, superlattices and harmonical confinement, for strongly correlated systems with $U=8$, which represents a challenge to DFT calculations. In general one finds that the deviations for the density profile are greater than the deviations for the ground-state energy and entanglement for all the systems. We attribute this to error cancellations related to the under and overestimate of DFT for the density profile: while Eq.(\ref{d_n}) will always sum up, thus producing a higher average deviation; Eqs. (\ref{d_e}) and (\ref{d_S}) will have smaller and greater contributions, thus on average producing a small deviation. While for the homogeneous systems there is a clear hierarchy between the deviations,  $D\%(\bar S)<D\%(e_0)<\bar D\%(n)$, and all the deviations decrease with $L$; for superlattices and harmonical confinement the relation among deviations is less trivial and strongly dependent on the SL structure and the confinement strength considered. For the superlattices, in general DFT calculations becomes less precise by increasing the number of impurities in the unit cell ($3:6$ and $4:5$). For the confined chains, DFT performs better for the metallic phases (i) ($k=0.002$) and (iii) ($k=0.015$). The highest deviations appear for the band-insulator phase (iv) ($k=0.035$). At this (iv) phase the density profile shows essentially only three values, $n_i=0$ at the chain borders, $n_i=1$ at the wings of the potential and $n_i=2$ at the core (see Fig. \ref{fig10}a), thus any imprecision on the DFT calculations has a bigger impact than in a metallic phase whose density profile is more homogeneously distributed. 

\section{Conclusions}

We have performed a comparative analysis between DFT and DMRG calculations, focusing on their efficacy in describing the electronic properties and quantum phase transitions of one-dimensional homogenous, superlattices, and harmonically confined nanostructures.  

For homogeneous finite chains, DFT exhibited significant deviations from the DMRG results for chain sizes $L\lesssim 60$, especially for the ground-state energy near the half-filling case: $\sim 16\%$ for small chains $L=10$, although it is less than 1\% for $L\sim 100$.  This arises from additional electronic correlations due to Friedel oscillations, which are not very well described by the local density approximation within DFT.  However, far from the half-filling case, around 20\% band filling, the DFT exhibited the best performance, presenting a maximum deviation of less than 2\% for the ground-state energy for any interaction and chain size. Also, there is a clear hierarchy between the deviations,  $D\%(\bar S)<D\%(e_0)< \bar D\%(n)$. 

For superlattice and confined systems the relation among the deviations is less trivial $-$ strongly dependent on the superlattice structure and the confinement strength $-$ and the inaccuracy of the DFT calculation is amplified, what is justified by the heterogeneous nature of the spatial distribution of electronic occupancy, representing a further challenge to the LDA approach. For the superlattices, in general increasing the number of impurities in the unit cell represents less precision of the DFT calculations. For the confined chains, DFT performs better for the metallic phases, while the highest deviations appear for the Mott and band-insulator phases.

 Although the DRMG offers an accurate description of the wave function of 1D systems, it faces computational constraints that impede its scalability to larger nanostructures: the DRMG calculation is restricted to the growth of the system entanglement, which depends on the chain length, while the DFT approach is practicable in arbitrary large systems. Furthermore, DFT results are generally obtained in shorter computational simulation times, typically on the order of minutes, while accurate convergence of the DMRG calculation can reach hours.\\

This comparative study highlights the importance of leveraging strategies that combine the capabilities of DFT and DMRG to predict electronic properties within one-dimensional nanostructures. Future research efforts can focus on developing
hybrid methodologies that synergistically employ the advantages of these approaches, especially in describing the correlation potential~\cite{10.1063/1.5129672}, thus offering a more robust and accurate framework for modeling and understanding nanoscale systems.

\bmhead{Acknowledgments}
This research was supported by FAPESP (2021/06744-8; 2023/00510-0; 2023/02293-7; 2021/02342-2), CNPq (403890/2021-7; 140854/2021-5;  306301/2022-9), Coordena\c{c}\~{a}o de Aperfei\c{c}oamento de Pessoal de Nivel Superior - Brasil (CAPES) - Finance Code 001, and by resources supplied by the Center for Scientific Computing (NCC/GridUNESP) from S\~{a}o Paulo State University.

\bibliography{bibliography}

\begin{thebibliography}{34}
\ifx \bisbn   \undefined \def \bisbn  #1{ISBN #1}\fi
\ifx \binits  \undefined \def \binits#1{#1}\fi
\ifx \bauthor  \undefined \def \bauthor#1{#1}\fi
\ifx \batitle  \undefined \def \batitle#1{#1}\fi
\ifx \bjtitle  \undefined \def \bjtitle#1{#1}\fi
\ifx \bvolume  \undefined \def \bvolume#1{\textbf{#1}}\fi
\ifx \byear  \undefined \def \byear#1{#1}\fi
\ifx \bissue  \undefined \def \bissue#1{#1}\fi
\ifx \bfpage  \undefined \def \bfpage#1{#1}\fi
\ifx \blpage  \undefined \def \blpage #1{#1}\fi
\ifx \burl  \undefined \def \burl#1{\textsf{#1}}\fi
\ifx \doiurl  \undefined \def \doiurl#1{\url{https://doi.org/#1}}\fi
\ifx \betal  \undefined \def \betal{\textit{et al.}}\fi
\ifx \binstitute  \undefined \def \binstitute#1{#1}\fi
\ifx \binstitutionaled  \undefined \def \binstitutionaled#1{#1}\fi
\ifx \bctitle  \undefined \def \bctitle#1{#1}\fi
\ifx \beditor  \undefined \def \beditor#1{#1}\fi
\ifx \bpublisher  \undefined \def \bpublisher#1{#1}\fi
\ifx \bbtitle  \undefined \def \bbtitle#1{#1}\fi
\ifx \bedition  \undefined \def \bedition#1{#1}\fi
\ifx \bseriesno  \undefined \def \bseriesno#1{#1}\fi
\ifx \blocation  \undefined \def \blocation#1{#1}\fi
\ifx \bsertitle  \undefined \def \bsertitle#1{#1}\fi
\ifx \bsnm \undefined \def \bsnm#1{#1}\fi
\ifx \bsuffix \undefined \def \bsuffix#1{#1}\fi
\ifx \bparticle \undefined \def \bparticle#1{#1}\fi
\ifx \barticle \undefined \def \barticle#1{#1}\fi
\bibcommenthead
\ifx \bconfdate \undefined \def \bconfdate #1{#1}\fi
\ifx \botherref \undefined \def \botherref #1{#1}\fi
\ifx \url \undefined \def \url#1{\textsf{#1}}\fi
\ifx \bchapter \undefined \def \bchapter#1{#1}\fi
\ifx \bbook \undefined \def \bbook#1{#1}\fi
\ifx \bcomment \undefined \def \bcomment#1{#1}\fi
\ifx \oauthor \undefined \def \oauthor#1{#1}\fi
\ifx \citeauthoryear \undefined \def \citeauthoryear#1{#1}\fi
\ifx \endbibitem  \undefined \def \endbibitem {}\fi
\ifx \bconflocation  \undefined \def \bconflocation#1{#1}\fi
\ifx \arxivurl  \undefined \def \arxivurl#1{\textsf{#1}}\fi
\csname PreBibitemsHook\endcsname

\bibitem[\protect\citeauthoryear{Geerlings
  et~al.}{2003}]{geerlings2003conceptual}
\begin{barticle}
\bauthor{\bsnm{Geerlings}, \binits{P.}},
\bauthor{\bsnm{De~Proft}, \binits{F.}},
\bauthor{\bsnm{Langenaeker}, \binits{W.}}:
\batitle{Conceptual density functional theory}.
\bjtitle{Chemical reviews}
\bvolume{103}(\bissue{5}),
\bfpage{1793}--\blpage{1874}
(\byear{2003})
\doiurl{10.1021/cr990029p}
\end{barticle}
\endbibitem

\bibitem[\protect\citeauthoryear{Cohen et~al.}{2012}]{cohen2012challenges}
\begin{barticle}
\bauthor{\bsnm{Cohen}, \binits{A.J.}},
\bauthor{\bsnm{Mori-S{\'a}nchez}, \binits{P.}},
\bauthor{\bsnm{Yang}, \binits{W.}}:
\batitle{Challenges for density functional theory}.
\bjtitle{Chemical reviews}
\bvolume{112}(\bissue{1}),
\bfpage{289}--\blpage{320}
(\byear{2012})
\doiurl{10.1021/cr200107z}
\end{barticle}
\endbibitem

\bibitem[\protect\citeauthoryear{Schollw{\"o}ck}{2011}]{dmrg2}
\begin{barticle}
\bauthor{\bsnm{Schollw{\"o}ck}, \binits{U.}}:
\batitle{The density-matrix renormalization group in the age of matrix product
  states}.
\bjtitle{Ann. Phys.}
\bvolume{326}(\bissue{1}),
\bfpage{96}--\blpage{192}
(\byear{2011})
\doiurl{10.1016/j.aop.2010.09.012}
\end{barticle}
\endbibitem

\bibitem[\protect\citeauthoryear{Kohn and Sham}{1965}]{PhysRev.140.A1133}
\begin{barticle}
\bauthor{\bsnm{Kohn}, \binits{W.}},
\bauthor{\bsnm{Sham}, \binits{L.J.}}:
\batitle{Self-consistent equations including exchange and correlation effects}.
\bjtitle{Phys. Rev.}
\bvolume{140},
\bfpage{1133}--\blpage{1138}
(\byear{1965})
\doiurl{10.1103/PhysRev.140.A1133}
\end{barticle}
\endbibitem

\bibitem[\protect\citeauthoryear{Burke}{2012}]{dft1}
\begin{barticle}
\bauthor{\bsnm{Burke}, \binits{K.}}:
\batitle{{Perspective on density functional theory}}.
\bjtitle{The Journal of Chemical Physics}
\bvolume{136}(\bissue{15}),
\bfpage{150901}
(\byear{2012})
\doiurl{10.1063/1.4704546}
{\href{https://arxiv.org/abs/https://pubs.aip.org/aip/jcp/article-pdf/doi/10.1063/1.4704546/14117227/150901\_1\_online.pdf}{{https://pubs.aip.org/aip/jcp/article-pdf/doi/10.1063/1.4704546/14117227/150901\_1\_online.pdf}}}
\end{barticle}
\endbibitem

\bibitem[\protect\citeauthoryear{Mirjani and Thijssen}{2011}]{ref1_vivian}
\begin{barticle}
\bauthor{\bsnm{Mirjani}, \binits{F.}},
\bauthor{\bsnm{Thijssen}, \binits{J.M.}}:
\batitle{Density functional theory based many-body analysis of electron
  transport through molecules}.
\bjtitle{Phys. Rev. B}
\bvolume{83},
\bfpage{035415}
(\byear{2011})
\doiurl{10.1103/PhysRevB.83.035415}
\end{barticle}
\endbibitem

\bibitem[\protect\citeauthoryear{Abedinpour et~al.}{2007}]{ref2_vivian}
\begin{barticle}
\bauthor{\bsnm{Abedinpour}, \binits{S.H.}},
\bauthor{\bsnm{Bakhtiari}, \binits{M.R.}},
\bauthor{\bsnm{Xianlong}, \binits{G.}},
\bauthor{\bsnm{Polini}, \binits{M.}},
\bauthor{\bsnm{Rizzi}, \binits{M.}},
\bauthor{\bsnm{Tosi}, \binits{M.P.}}:
\batitle{Phase behaviors of strongly correlated fermi gases in one-dimensional
  confinements}.
\bjtitle{Laser Physics}
\bvolume{17}(\bissue{2}),
\bfpage{162}--\blpage{168}
(\byear{2007})
\doiurl{10.1134/S1054660X0702020X}
\end{barticle}
\endbibitem

\bibitem[\protect\citeauthoryear{Hu et~al.}{2010}]{ref3_vivian}
\begin{barticle}
\bauthor{\bsnm{Hu}, \binits{J.-H.}},
\bauthor{\bsnm{Wang}, \binits{J.-J.}},
\bauthor{\bsnm{Xianlong}, \binits{G.}},
\bauthor{\bsnm{Okumura}, \binits{M.}},
\bauthor{\bsnm{Igarashi}, \binits{R.}},
\bauthor{\bsnm{Yamada}, \binits{S.}},
\bauthor{\bsnm{Machida}, \binits{M.}}:
\batitle{Ground-state properties of the one-dimensional attractive hubbard
  model with confinement: A comparative study}.
\bjtitle{Phys. Rev. B}
\bvolume{82},
\bfpage{014202}
(\byear{2010})
\doiurl{10.1103/PhysRevB.82.014202}
\end{barticle}
\endbibitem

\bibitem[\protect\citeauthoryear{Sauban\`ere and Pastor}{2011}]{ref5_vivian}
\begin{barticle}
\bauthor{\bsnm{Sauban\`ere}, \binits{M.}},
\bauthor{\bsnm{Pastor}, \binits{G.M.}}:
\batitle{Density-matrix functional study of the hubbard model on one- and
  two-dimensional bipartite lattices}.
\bjtitle{Phys. Rev. B}
\bvolume{84},
\bfpage{035111}
(\byear{2011})
\doiurl{10.1103/PhysRevB.84.035111}
\end{barticle}
\endbibitem

\bibitem[\protect\citeauthoryear{Akande and Sanvito}{2010}]{ref4_vivian}
\begin{barticle}
\bauthor{\bsnm{Akande}, \binits{A.}},
\bauthor{\bsnm{Sanvito}, \binits{S.}}:
\batitle{Electric field response of strongly correlated one-dimensional metals:
  A bethe ansatz density functional theory study}.
\bjtitle{Phys. Rev. B}
\bvolume{82},
\bfpage{245114}
(\byear{2010})
\doiurl{10.1103/PhysRevB.82.245114}
\end{barticle}
\endbibitem

\bibitem[\protect\citeauthoryear{Giamarchi}{2003}]{giamarchi2003quantum}
\begin{bbook}
\bauthor{\bsnm{Giamarchi}, \binits{T.}}:
\bbtitle{Quantum Physics in One Dimension}
vol. \bseriesno{121}.
\bpublisher{Clarendon Press},
\blocation{Oxford}
(\byear{2003})
\end{bbook}
\endbibitem

\bibitem[\protect\citeauthoryear{Mendes-Santos
  et~al.}{2013}]{PhysRevB.87.214407}
\begin{barticle}
\bauthor{\bsnm{Mendes-Santos}, \binits{T.}},
\bauthor{\bsnm{Paiva}, \binits{T.}},
\bauthor{\bsnm{Santos}, \binits{R.R.}}:
\batitle{Entanglement, magnetism, and metal-insulator transitions in fermionic
  superlattices}.
\bjtitle{Phys. Rev. B}
\bvolume{87},
\bfpage{214407}
(\byear{2013})
\doiurl{10.1103/PhysRevB.87.214407}
\end{barticle}
\endbibitem

\bibitem[\protect\citeauthoryear{Paiva and dos Santos}{1998}]{PhysRevB.58.9607}
\begin{barticle}
\bauthor{\bsnm{Paiva}, \binits{T.}},
\bauthor{\bsnm{Santos}, \binits{R.R.}}:
\batitle{Metal-insulator transition in one-dimensional hubbard superlattices}.
\bjtitle{Phys. Rev. B}
\bvolume{58},
\bfpage{9607}--\blpage{9610}
(\byear{1998})
\doiurl{10.1103/PhysRevB.58.9607}
\end{barticle}
\endbibitem

\bibitem[\protect\citeauthoryear{Liu et~al.}{2020}]{PhysRevB.102.235151}
\begin{barticle}
\bauthor{\bsnm{Liu}, \binits{T.}},
\bauthor{\bsnm{He}, \binits{J.J.}},
\bauthor{\bsnm{Yoshida}, \binits{T.}},
\bauthor{\bsnm{Xiang}, \binits{Z.-L.}},
\bauthor{\bsnm{Nori}, \binits{F.}}:
\batitle{Non-hermitian topological mott insulators in one-dimensional fermionic
  superlattices}.
\bjtitle{Phys. Rev. B}
\bvolume{102},
\bfpage{235151}
(\byear{2020})
\doiurl{10.1103/PhysRevB.102.235151}
\end{barticle}
\endbibitem

\bibitem[\protect\citeauthoryear{Zanardi}{2002}]{zanardi}
\begin{barticle}
\bauthor{\bsnm{Zanardi}, \binits{P.}}:
\batitle{Quantum entanglement in fermionic lattices}.
\bjtitle{Phys. Rev. A}
\bvolume{65},
\bfpage{042101}
(\byear{2002})
\doiurl{10.1103/PhysRevA.65.042101}
\end{barticle}
\endbibitem

\bibitem[\protect\citeauthoryear{Tichy et~al.}{2013}]{tichy2013entanglement}
\begin{barticle}
\bauthor{\bsnm{Tichy}, \binits{M.C.}},
\bauthor{\bsnm{Melo}, \binits{F.}},
\bauthor{\bsnm{Ku{\'s}}, \binits{M.}},
\bauthor{\bsnm{Mintert}, \binits{F.}},
\bauthor{\bsnm{Buchleitner}, \binits{A.}}:
\batitle{Entanglement of identical particles and the detection process}.
\bjtitle{Fortschritte der Physik}
\bvolume{61}(\bissue{2-3}),
\bfpage{225}--\blpage{237}
(\byear{2013})
\doiurl{10.1002/prop.201200079}
\end{barticle}
\endbibitem

\bibitem[\protect\citeauthoryear{Canella and Fran\c{c}a}{2021}]{canellaVV}
\begin{barticle}
\bauthor{\bsnm{Canella}, \binits{G.A.}},
\bauthor{\bsnm{Fran\c{c}a}, \binits{V.V.}}:
\batitle{Mott-anderson metal-insulator transitions from entanglement}.
\bjtitle{Phys. Rev. B}
\bvolume{104},
\bfpage{134201}
(\byear{2021})
\doiurl{10.1103/PhysRevB.104.134201}
\end{barticle}
\endbibitem

\bibitem[\protect\citeauthoryear{Fishman et~al.}{2022}]{itensor}
\begin{botherref}
\oauthor{\bsnm{Fishman}, \binits{M.}},
\oauthor{\bsnm{White}, \binits{S.R.}},
\oauthor{\bsnm{Stoudenmire}, \binits{E.M.}}:
{The ITensor Software Library for Tensor Network Calculations}.
SciPost Phys. Codebases,
4
(2022)
\doiurl{10.21468/SciPostPhysCodeb.4}
\end{botherref}
\endbibitem

\bibitem[\protect\citeauthoryear{Capelle and Campo}{2013}]{CAPELLE201391}
\begin{barticle}
\bauthor{\bsnm{Capelle}, \binits{K.}},
\bauthor{\bsnm{Campo}, \binits{V.L.}}:
\batitle{Density functionals and model hamiltonians: Pillars of many-particle
  physics}.
\bjtitle{Physics Reports}
\bvolume{528}(\bissue{3}),
\bfpage{91}--\blpage{159}
(\byear{2013})
\doiurl{10.1016/j.physrep.2013.03.002} .
\bcomment{Density functionals and model Hamiltonians: Pillars of many-particle
  physics}
\end{barticle}
\endbibitem

\bibitem[\protect\citeauthoryear{França et~al.}{2012}]{FVC}
\begin{barticle}
\bauthor{\bsnm{França}, \binits{V.V.}},
\bauthor{\bsnm{Vieira}, \binits{D.}},
\bauthor{\bsnm{Capelle}, \binits{K.}}:
\batitle{Simple parameterization for the ground-state energy of the infinite
  hubbard chain incorporating mott physics, spin-dependent phenomena and
  spatial inhomogeneity}.
\bjtitle{New Journal of Physics}
\bvolume{14}(\bissue{7}),
\bfpage{073021}
(\byear{2012})
\doiurl{10.1088/1367-2630/14/7/073021}
\end{barticle}
\endbibitem

\bibitem[\protect\citeauthoryear{Lieb and Wu}{1968}]{PhysRevLett.20.1445}
\begin{barticle}
\bauthor{\bsnm{Lieb}, \binits{E.H.}},
\bauthor{\bsnm{Wu}, \binits{F.Y.}}:
\batitle{Absence of mott transition in an exact solution of the short-range,
  one-band model in one dimension}.
\bjtitle{Phys. Rev. Lett.}
\bvolume{20},
\bfpage{1445}--\blpage{1448}
(\byear{1968})
\doiurl{10.1103/PhysRevLett.20.1445}
\end{barticle}
\endbibitem

\bibitem[\protect\citeauthoryear{Essler et~al.}{2005}]{essler2005one}
\begin{bbook}
\bauthor{\bsnm{Essler}, \binits{F.H.}},
\bauthor{\bsnm{Frahm}, \binits{H.}},
\bauthor{\bsnm{G{\"o}hmann}, \binits{F.}},
\bauthor{\bsnm{Kl{\"u}mper}, \binits{A.}},
\bauthor{\bsnm{Korepin}, \binits{V.E.}}:
\bbtitle{The One-dimensional Hubbard Model}.
\bpublisher{Cambridge University Press},
\blocation{New York}
(\byear{2005})
\end{bbook}
\endbibitem

\bibitem[\protect\citeauthoryear{Bed\"urftig et~al.}{1998}]{PhysRevB.58.10225}
\begin{barticle}
\bauthor{\bsnm{Bed\"urftig}, \binits{G.}},
\bauthor{\bsnm{Brendel}, \binits{B.}},
\bauthor{\bsnm{Frahm}, \binits{H.}},
\bauthor{\bsnm{Noack}, \binits{R.M.}}:
\batitle{Friedel oscillations in the open hubbard chain}.
\bjtitle{Phys. Rev. B}
\bvolume{58},
\bfpage{10225}--\blpage{10235}
(\byear{1998})
\doiurl{10.1103/PhysRevB.58.10225}
\end{barticle}
\endbibitem

\bibitem[\protect\citeauthoryear{S\"offing et~al.}{2009}]{PhysRevB.79.195114}
\begin{barticle}
\bauthor{\bsnm{S\"offing}, \binits{S.A.}},
\bauthor{\bsnm{Bortz}, \binits{M.}},
\bauthor{\bsnm{Schneider}, \binits{I.}},
\bauthor{\bsnm{Struck}, \binits{A.}},
\bauthor{\bsnm{Fleischhauer}, \binits{M.}},
\bauthor{\bsnm{Eggert}, \binits{S.}}:
\batitle{Wigner crystal versus friedel oscillations in the one-dimensional
  hubbard model}.
\bjtitle{Phys. Rev. B}
\bvolume{79},
\bfpage{195114}
(\byear{2009})
\doiurl{10.1103/PhysRevB.79.195114}
\end{barticle}
\endbibitem

\bibitem[\protect\citeauthoryear{Arisa and
  Fran\c{c}a}{2020}]{PhysRevB.101.214522}
\begin{barticle}
\bauthor{\bsnm{Arisa}, \binits{D.}},
\bauthor{\bsnm{Fran\c{c}a}, \binits{V.V.}}:
\batitle{Linear mapping between magnetic susceptibility and entanglement in
  conventional and exotic one-dimensional superfluids}.
\bjtitle{Phys. Rev. B}
\bvolume{101},
\bfpage{214522}
(\byear{2020})
\doiurl{10.1103/PhysRevB.101.214522}
\end{barticle}
\endbibitem

\bibitem[\protect\citeauthoryear{Br\"unner et~al.}{2013}]{PhysRevA.87.032311}
\begin{barticle}
\bauthor{\bsnm{Br\"unner}, \binits{T.}},
\bauthor{\bsnm{Runge}, \binits{E.}},
\bauthor{\bsnm{Buchleitner}, \binits{A.}},
\bauthor{\bsnm{Fran\c{c}a}, \binits{V.V.}}:
\batitle{Entanglement enhancement in spatially inhomogeneous many-body
  systems}.
\bjtitle{Phys. Rev. A}
\bvolume{87},
\bfpage{032311}
(\byear{2013})
\doiurl{10.1103/PhysRevA.87.032311}
\end{barticle}
\endbibitem

\bibitem[\protect\citeauthoryear{Fran\c{c}a et~al.}{2012}]{PhysRevA.86.033622}
\begin{barticle}
\bauthor{\bsnm{Fran\c{c}a}, \binits{V.V.}},
\bauthor{\bsnm{H\"orndlein}, \binits{D.}},
\bauthor{\bsnm{Buchleitner}, \binits{A.}}:
\batitle{Fulde-ferrell-larkin-ovchinnikov critical polarization in
  one-dimensional fermionic optical lattices}.
\bjtitle{Phys. Rev. A}
\bvolume{86},
\bfpage{033622}
(\byear{2012})
\doiurl{10.1103/PhysRevA.86.033622}
\end{barticle}
\endbibitem

\bibitem[\protect\citeauthoryear{Coe et~al.}{2010}]{PhysRevA.81.052321}
\begin{barticle}
\bauthor{\bsnm{Coe}, \binits{J.P.}},
\bauthor{\bsnm{Fran\c{c}a}, \binits{V.V.}},
\bauthor{\bsnm{D'Amico}, \binits{I.}}:
\batitle{Hubbard model as an approximation to the entanglement in
  nanostructures}.
\bjtitle{Phys. Rev. A}
\bvolume{81},
\bfpage{052321}
(\byear{2010})
\doiurl{10.1103/PhysRevA.81.052321}
\end{barticle}
\endbibitem

\bibitem[\protect\citeauthoryear{Larsson and
  Johannesson}{2006}]{PhysRevA.73.042320}
\begin{barticle}
\bauthor{\bsnm{Larsson}, \binits{D.}},
\bauthor{\bsnm{Johannesson}, \binits{H.}}:
\batitle{Single-site entanglement of fermions at a quantum phase transition}.
\bjtitle{Phys. Rev. A}
\bvolume{73},
\bfpage{042320}
(\byear{2006})
\doiurl{10.1103/PhysRevA.73.042320}
\end{barticle}
\endbibitem

\bibitem[\protect\citeauthoryear{Anfossi et~al.}{2006}]{PhysRevB.73.085113}
\begin{barticle}
\bauthor{\bsnm{Anfossi}, \binits{A.}},
\bauthor{\bsnm{Boschi}, \binits{C.D.E.}},
\bauthor{\bsnm{Montorsi}, \binits{A.}},
\bauthor{\bsnm{Ortolani}, \binits{F.}}:
\batitle{Single-site entanglement at the superconductor-insulator transition in
  the hirsch model}.
\bjtitle{Phys. Rev. B}
\bvolume{73},
\bfpage{085113}
(\byear{2006})
\doiurl{10.1103/PhysRevB.73.085113}
\end{barticle}
\endbibitem

\bibitem[\protect\citeauthoryear{Osborne and
  Nielsen}{2002}]{PhysRevA.66.032110}
\begin{barticle}
\bauthor{\bsnm{Osborne}, \binits{T.J.}},
\bauthor{\bsnm{Nielsen}, \binits{M.A.}}:
\batitle{Entanglement in a simple quantum phase transition}.
\bjtitle{Phys. Rev. A}
\bvolume{66},
\bfpage{032110}
(\byear{2002})
\doiurl{10.1103/PhysRevA.66.032110}
\end{barticle}
\endbibitem

\bibitem[\protect\citeauthoryear{Paviglianiti and
  Silva}{2023}]{PhysRevB.108.184302}
\begin{barticle}
\bauthor{\bsnm{Paviglianiti}, \binits{A.}},
\bauthor{\bsnm{Silva}, \binits{A.}}:
\batitle{Multipartite entanglement in the measurement-induced phase transition
  of the quantum ising chain}.
\bjtitle{Phys. Rev. B}
\bvolume{108},
\bfpage{184302}
(\byear{2023})
\doiurl{10.1103/PhysRevB.108.184302}
\end{barticle}
\endbibitem

\bibitem[\protect\citeauthoryear{Fran\c{c}a and
  Capelle}{2006}]{PhysRevA.74.042325}
\begin{barticle}
\bauthor{\bsnm{Fran\c{c}a}, \binits{V.V.}},
\bauthor{\bsnm{Capelle}, \binits{K.}}:
\batitle{Entanglement of strongly interacting low-dimensional fermions in
  metallic, superfluid, and antiferromagnetic insulating systems}.
\bjtitle{Phys. Rev. A}
\bvolume{74},
\bfpage{042325}
(\byear{2006})
\doiurl{10.1103/PhysRevA.74.042325}
\end{barticle}
\endbibitem

\bibitem[\protect\citeauthoryear{Baiardi and Reiher}{2020}]{10.1063/1.5129672}
\begin{barticle}
\bauthor{\bsnm{Baiardi}, \binits{A.}},
\bauthor{\bsnm{Reiher}, \binits{M.}}:
\batitle{{The density matrix renormalization group in chemistry and molecular
  physics: Recent developments and new challenges}}.
\bjtitle{The Journal of Chemical Physics}
\bvolume{152}(\bissue{4}),
\bfpage{040903}
(\byear{2020})
\doiurl{10.1063/1.5129672}
{\href{https://arxiv.org/abs/https://pubs.aip.org/aip/jcp/article-pdf/doi/10.1063/1.5129672/13277206/040903\_1\_online.pdf}{{https://pubs.aip.org/aip/jcp/article-pdf/doi/10.1063/1.5129672/13277206/040903\_1\_online.pdf}}}
\end{barticle}
\endbibitem

\end{thebibliography}


\end{document}